\documentclass[article,shortnames]{jss}
\usepackage{booktabs}
\usepackage{bm} 

\providecommand{\binomdistn}{\mathrm{Binomial}}
\providecommand{\normaldistn}{\mathrm{Normal}}
\providecommand{\logit}{\mathrm{logit}}
\providecommand{\Emax}{\mathrm{E}_\mathrm{max}}
\providecommand{\EDfifty}{\mathrm{ED}_{50}}

\author{Burak K\"{u}rsad G\"{u}nhan\\ Merck KGaA, Darmstadt, Germany
        \And Christian R\"{o}ver\\ University Medical Center G\"{o}ttingen
        \AND Tim Friede\\ University Medical Center G\"{o}ttingen}
\title{\pkg{MetaStan}: An \proglang{R} package for Bayesian (model-based) meta-analysis using \proglang{Stan}}

\Plainauthor{Burak K\"{u}rsad G\"{u}nhan, Christian R\"{o}ver, Tim Friede}
\Plaintitle{MetaStan: An R package for Bayesian (model-based) meta-analysis using Stan}

\Abstract{
  Meta-analysis methods are used to combine evidence from multiple studies. Meta-regression as well as model-based meta-analysis are extensions of standard pairwise meta-analysis in which information about study-level covariates and (arm-level) dosing amount or exposure may be taken into account. A Bayesian approach to inference is very attractive in this context, especially when a meta-analysis is based on few studies only or rare events. In this article, we present the \proglang{R}~package \pkg{MetaStan} which implements a wide range of pairwise and model-based meta-analysis models. 

  A generalised linear mixed model (GLMM) framework is used to describe the pairwise meta-analysis, meta-regression and model-based meta-analysis models. Within the GLMM framework, the likelihood and link functions are adapted to reflect the nature of the data. For example, a binomial likelihood with a logit link is used to perform a meta-analysis based on datasets with dichotomous endpoints. Bayesian computations are conducted using \proglang{Stan} via the \pkg{rstan} interface. \proglang{Stan} uses a Hamiltonian Monte Carlo sampler which belongs to the family of Markov chain Monte Carlo methods. \proglang{Stan} implementations are done by using suitable parametrizations to ease computations.

  The user-friendly \proglang{R}~package \pkg{MetaStan}, available on CRAN, supports a wide range of pairwise and model-based meta-analysis models. \pkg{MetaStan} provides fitting functions for pairwise meta-analysis with the option of including covariates and model-based meta-analysis. The supported outcome types are continuous, binary, and count. Forest plots for the pairwise meta-analysis and dose-response plots for the model-based meta-analysis can be obtained from the package. The use of \pkg{MetaStan} is demonstrated through clinical examples.
}

\Plainkeywords{random-effects meta-analysis, meta-regression, dose-response curve}

\Address{
  Burak K.\ G\"{u}nhan\\
  Merck KGaA\\
  Frankfurter Straße 250\\
  64293 Darmstadt, Germany\\
  E-mail: \email{burakgunhan@gmail.com}\\
  ORCID iD: \href{http://orcid.org/0000-0002-7454-8680}{0000-0002-7454-8680}
  
  Christian Röver\\
  University Medical Center Göttingen, Dept. Medical Statistics\\
  Humboldtallee 32\\
  37073 Göttingen, Germany\\
  E-mail: \email{christian.roever@med.uni-goettingen.de}\\
  ORCID iD: \href{https://orcid.org/0000-0002-6911-698X}{0000-0002-6911-698X}

  Tim Friede\\
  University Medical Center Göttingen, Dept. Medical Statistics\\
  Humboldtallee 32\\
  37073 Göttingen, Germany\\
  E-mail: \email{tim.friede@med.uni-goettingen.de}\\
  ORCID iD: \href{https://orcid.org/0000-0001-5347-7441}{0000-0001-5347-7441}
}

\begin{document}


\section{Introduction}\label{sec:intro}
Meta-analyses are used to combine evidence from multiple clinical trials in order to address a question of interest, for example, to estimate the efficacy or the safety profile of a drug or other therapy. The consideration of potential heterogeneity (variability) between trials is an important aspect of a meta-analysis. Thus, statistical methods which model the heterogeneity between trials, so-called \emph{random-effects models}, have been suggested for the meta-analysis of clinical trials \citep{Higgins2009}. Conventionally, pairwise comparisons (``treatment'' vs ``control'' arm) are investigated in a standard \emph{pairwise} meta-analysis. \emph{Network} meta-analysis (NMA) \citep{Lumley2002} is a generalization of a pairwise meta-analysis, in which multiple treatments are compared. Furthermore, pairwise meta-analysis approaches are readily generalized to \emph{meta-regression} models in order to account for trial-level covariates, for example, the randomization status of each trial. Such approaches can be used to explain potential sources of heterogeneity between trials \citep{Sutton2001,Thompson1994}.

In phase~II of clinical drug development, trials are conducted to investigate the dose-response relationship of a compound. Meta-analyses of dose-response trials are called \emph{model-based meta-analyses} (MBMA). MBMA can be seen as a generalization of pairwise meta-analysis, in which information about dosing amount or exposure for each study arm is taken into account \citep{Mandema2005}. In an MBMA, dose-response relationships are modeled using a variety of functional forms including Emax or linear models \citep{Mawdsley2016}. In particular, studies investigating multiple doses of the same compund within a single trial (multi-arm trials) can be included in the dataset. 

A Bayesian framework has been suggested and is in fact commonly used for pairwise and model-based meta-analysis \citep{Smith1995, Sutton2001, Mawdsley2016}. In a Bayesian framework, prior distributions need to be specified for all parameters in the model. Priors need to be chosen carefully, but the use of weakly informative priors may be particularly beneficial e.g.\ in cases of few studies or rare events \citep{RoeverEtAl2021}. Prior specification may also be thought of as related to regularisation or penalisation
approaches \citep{Greenland2015}. Long-run properties of the use of weakly informative priors for such meta-analysis scenarios were investigated and suggested by \citet{Friede2017} and \citet{Guenhan2020Meta}.

\proglang{R}~is a freely available programming language for statistical computing and visualization \citep{R}. There are several \proglang{R} packages implementing pairwise meta-analysis and meta-regression models in a Bayesian framework utilizing different computational methods. For instance, the \pkg{bayesmeta} package \citep{Roever2020bayesmeta} uses semi-analytic and semi-numeric methods, the \pkg{bmeta} package \citep{bmeta} uses a Metropolis-Hastings (Gibbs) sampler via \proglang{JAGS} \citep{Plummer2003}, and the \pkg{baggr} package \citep{baggr} uses Hamiltonian Monte Carlo via the \pkg{rstan} \proglang{R} package \citep{rstan}. The \pkg{MBNMAdose} package \citep{MBNMAdose} has been developed for model-based NMA using \proglang{JAGS} for computation. Although \pkg{MBNMAdose} can also be used for model-based meta-analysis of pairwise comparisons, the focus of \pkg{MBNMAdose} is on model-based network meta-analysis. Thus, the NMA terminology used in the \pkg{MBNMAdose} package might be a hurdle for practitioners interested in model-based meta-analysis of pairwise comparisons. Furthermore, it has been argued that the Hamiltonian Monte Carlo used by \proglang{Stan} is a more efficient and robust sampler than Gibbs or Metropolis-Hastings samplers used by \proglang{JAGS} for models with complex posterior distributions \citep{Stan}. 

The main contribution of this paper is to present the \proglang{R} package \pkg{MetaStan} \citep{MetaStan}, which provides a  Bayesian implementation of a wide range of pairwise meta-analysis, meta-regression, and MBMA methods for pairwise comparisons. The \pkg{MetaStan} package relies on the \pkg{rstan} package for computation to obtain posterior distributions. Statistical analysis is implemented via a generalized linear mixed model (GLMM) approach, allowing the user to specify suitable likelihood and link functions reflecting the nature of the data. The covered analysis models fall into the class of \emph{one-stage} models, which, unlike some \emph{two-stage} methods, do not need to rely on a normal approximation involving intermediately computed ``effect sizes'' and associated standard errors, which is sometimes problematic \citep{Burke2017, Jackson2018}.

The paper is structured as follows. The underlying statistical models of the \proglang{R} package \pkg{Meta\-Stan} are described in Section~\ref{sec:models}. Section~\ref{sec:stan} discusses the model implementation using \proglang{Stan}. The main functions of the package and some practical data applications from clinical medicine are discussed in Section~\ref{sec:usage}. Section~\ref{sec:discuss} provides some conclusions.

\section{Statistical models} \label{sec:models}
\subsection{Preliminary remark}
We describe the underlying theory by using a generalized linear mixed model (GLMM) framework. In Section~\ref{sec:PMA}, we describe a pairwise meta-analysis model, then a meta-regression model in Section~\ref{sec:MReg}, and we continue with an MBMA in Section~\ref{sec:MBMA}.

\subsection{Pairwise meta-analysis} \label{sec:PMA}
Pairwise meta-analysis for binary outcomes has been introduced by \citet{Smith1995} and also discussed by \citet{Guenhan2020Meta} among others. Here we consider the general case of datasets with different types of outcomes, namely binary, continuous, or count. 
We specify a model for the data~$y_{ij}$ from the $j$th~arm of the $i$th~study (where $i \in \{1,\ldots,k\}$ and $j \in \{0,1\}$).
The observables~$y_{ij}$ here may be Bernoulli or Poisson counts, or continuous outcomes modelled via a normal model.
The distributions of outcomes in the two arms are assumed to differ by a parameter~$\theta$, e.g., an event probability, an event rate, or a mean.
A link function $G(\cdot)$ is used to map the parameter~$\theta$ to a scale where effects are assumed additive; the parameter for the $j$th arm in the $i$th study then is defined through
\begin{equation}
      G(\theta_{ij}) \;=\;  \left\{
      \begin{array}{ll}
        \mu_{i} - 0.5 (d + \gamma_{i}) & \mbox{for $j=0$ (control arm)} \\
        \mu_{i} + 0.5 (d + \gamma_{i}) & \mbox{for $j=1$ (treatment arm),} \label{eq:PMA}
      \end{array} \right.
\end{equation}
where the $\mu_{i}$ are fixed effects denoting a baseline in trial~$i$, $d$~is the mean treatment effect, 
and $d_i=d + \gamma_i$ is the (study specific) ``random'' treatment effect in the $i$th study.
Depending on the type of endpoint (binary, continuous or count), the~$\mu_i$ relate to baseline probabilities, levels or rates, and effectively facilitate a stratification by study.
Random effects are included to reflect heterogeneity and they are assummed to be normally distributed, $\gamma_{i}  \sim \normaldistn(0, \tau^2)$. The standard deviation~$\tau$ is a measure of the degree of heterogeneity between trials. When $\tau = 0$ (implying $\gamma_{i} = 0$), the model reduces to a common-effect model. By treating baselines~$\mu_{i}$ as fixed effects, we are focusing on \emph{relative} treatment effects, for example, log odds ratios for binary data.

A common variation of the parametrization in (\ref{eq:PMA}) is obtained by using a different coding, namely 
\begin{equation}
      G(\theta_{ij}) \;=\;  \left\{
      \begin{array}{ll}
        \mu_{i} & \mbox{for $j=0$ (control arm)} \\
        \mu_{i} + d + \gamma_{i} & \mbox{for $j=1$ (treatment arm).} \label{eq:PMA2}
      \end{array} \right.
\end{equation}
In \pkg{MetaStan}, the model with treatment coding as in~(\ref{eq:PMA}) is used by default due to symmetry arguments, and since it has been shown to perform better in the terms of heterogeneity estimation \citep{Jackson2018}.
However, the model as in~(\ref{eq:PMA2}) is available as an option, and a variation of this latter model as will also be utilized for model-based meta-analysis later on (see Section~\ref{sec:MBMA} below).

A common aim of inference is the \emph{prediction} of a ``new'' (future) treatment effect; for a given heterogeneity~$\tau$, a prediction of treatment effect from a new study ($d^\star$) is defined by the conditional expression \begin{equation}
  d^\star|d,\tau \;\sim\; \normaldistn(d, \tau^2)\mbox{.}
\end{equation}
Now consider the above model for binary outcomes as an example. The event counts~$y_{ij}$ and the numbers of patients~($n_{ij}$) are given for the $i$th arm in the $j$th trial. The likelihood function can be expressed via $y_{ij} \sim \binomdistn(\theta_{ij},n_{ij})$, where a \emph{logit} link function is used to map the event probabilities~$\theta_{ij}$ to the log-odds scale ($G(\theta_{ij}) = \logit(\theta_{ij})$) in Equation~(\ref{eq:PMA}). The effect~$d$ then corresponds to a logarithmic odds ratio. Likewise, the models for continuous outcome data can be formulated using a normal likelihood with identity link function. When the data available for the meta-analysis are counts, a Poisson likelihood with log link can be used, and the effect~$\theta$ then corresponds to a logarithmic rate ratio. 

\subsection{Meta-regression} \label{sec:MReg}
Meta-regression is used to investigate associations between characteristics of trials and treatment effects \citep{LauIoannidisSchmid1998,Sutton2001,ThompsonHiggins2002,TiptonEtAl2019a,HigginsEtAl2021}, 
which technically constitute \emph{interactions} between study-level covariables and the effects of interest \citep{DoneganEtAl2015}.
Thus, meta-regression is helpful to explore potential sources of heterogeneity between trials \citep{Thompson1994,Higgins2008}. A meta-regression model can be constructed by including a vector of (one or several) trial-level covariates~$\mathbf{x}_{i}$, such as the mean age of the subjects in each trial. To generalize the model, Equation~(\ref{eq:PMA}) becomes
\begin{equation}
      G(\theta_{ij}) \;=\;  \left\{
      \begin{array}{ll}
        \mu_{i} - 0.5 ( d + \mathbf{x}_{i}^\prime \bm{\beta} +  \gamma_{i}) & \mbox{for $j=0$ (control arm)} \\
        \mu_{i} + 0.5 ( d + \mathbf{x}_{i}^\prime \bm{\beta} + \gamma_{i}) & \mbox{for $j=1$ (treatment arm),} \label{eq:MReg}
      \end{array} \right.
\end{equation}
where $\bm{\beta}$ is the vector of regression coefficients reflecting the influence of covariables on the (study-specific) treatment effect~$d_i$. The original model~(\ref{eq:PMA}) again constitutes the special case of only fitting a single ``intercept'' coefficient. Note that $\mathbf{x}_{i}^\prime \beta$ may also include interactions of covariates, which are not regularly considered in practice but may be important in some applications \citep[see e.g.][]{KnopEtAl2022}.

\subsection{Model-based meta-analysis} \label{sec:MBMA}
The MBMA model discussed in this section has been introduced by \citet{Mawdsley2016} for network meta-analysis. Here, we adapt their model for pairwise comparisons, that is,  only two treatments are compared.  Assume that a covariate~$\delta_{ij}$, which may often be thought of as a \emph{dose} or \emph{exposure} measure, is given for each dose group~$j$ of trial~$i$. A key part of the MBMA model is the functional form of the dose-response function. Different functional forms $f_{\bm{d}}(\delta_{ij})$, where $\bm{d}$ refers to the parameters defining the exact dose-response relationship \citep{Bretz2005}. Four common dose-response models, linear, linear log-dose, $\Emax$ and sigmoidal $\Emax$, are implemented in the \pkg{MetaStan} package and are detailed in Table~\ref{tab:doseresponse}. The parameter~$\Emax$ is the maximum effect attributable to the drug, the parameter~$\EDfifty$ is the dose at which half of the maximum effect is reached, and $n$~is the \emph{Hill parameter} controlling the shape of the dose-response curve.

\begin{table}[b]
\centering
\caption{\label{tab:doseresponse}Four commonly used dose-response models available in \pkg{MetaStan}. Also, the parameters are shortly explained in the table.}
\begin{tabular}{lll}
  \toprule
  model             &  dose-response $f_{\bm{d}}(\delta_{i,k})$  & parameter(s) $\bm{d}$ \\
  \midrule
  linear            &  $\alpha \cdot \delta_{i,k}$  &  $\alpha$ (slope) \\
  linear log-dose   &  $\alpha \cdot \log(\delta_{i,k} + 1)$ &  $\alpha$ (slope)\\ 
  $\Emax$           &  $\Emax \cdot \delta_{i,k}/(\EDfifty + \delta_{i,k})$  & $\Emax$ (max.\ effect),  $\EDfifty$ (a dose) \\ 
  sigmoidal $\Emax$ &  $\Emax \cdot \delta_{i,k}^n/(\EDfifty + \delta_{i,k}^n)$    & $\Emax$ (max.\ effect),  $\EDfifty$ (a dose), $n$ (shape) \\ 
  \bottomrule
\end{tabular}
\end{table}

In an MBMA model, the $\delta_{i,k}$ for a dose group~$j$ of trial~$i$ can be modeled similar to Equation~(\ref{eq:PMA2}), i.e.
\begin{equation}
      G(\theta_{ij}) \;=\; \left\{
      \begin{array}{ll}
        \mu_{i} & \mbox{for $j=0$ (control arm)} \\
        \mu_{i} + f(\delta_{ij}) + \gamma_{ij}  &\mbox{for $j\geq1$ (treatment arms).} \label{eq:MBMA}
      \end{array} \right.
\end{equation}
As in the model described in Section~\ref{sec:PMA}, we assume that baseline parameters~$\mu_i$ are fixed effects. The described MBMA model is a generalization of the pairwise meta-analysis from Equation~(\ref{eq:PMA2}). For two-armed trials, we assume that $\gamma_{ij} \sim \normaldistn(0, \tau^2)$. Unlike the pairwise meta-analysis, MBMA may include multi-arm trials, where all arms are modelled relative to the same reference group (placebo, or $\delta_i=0$). Thus, we can not assume independence of random-effects $\gamma_{ij}$ within the same trial. To account for this, we use a multivariate normal distribution
\begin{equation}
\bm{\gamma}_{i} \;\sim\; \normaldistn_{T - 1}(\bm{0}, \bm{\Sigma}_{\gamma}) \label{heterVar}
\end{equation}
where $T$ is the number of dose levels, and $\bm{\Sigma}_{\gamma}$ is a symmetric homogeneous covariance matrix with diagonal entries equal to $\tau^2$ and all off-diagonal entries set to $\tau^2/2$ \citep{Mawdsley2016}.

\subsection{Prior distributions} \label{sec:prior}
To fit the models in a Bayesian framework, prior distributions for all model parameters need to be specified. In the meta-analysis literature, it is common to use vague (or noninformative) priors for the baseline parameters~$\mu_{i}$, the treatment effect parameter~$\theta$, and the coefficient vector~$\bm{\beta}$ in the meta-regression model \citep{Sutton2001}. As a vague prior, a normal prior with a ``neutral'' mean and some ``large'' variance (such as $\normaldistn(0, 100^2)$) can be used.  For meta-analysis of few studies involving rare events, \citet{Guenhan2020Meta} suggested the use of weakly informative priors for the treatment effect parameter, which may also be seen as a form of penalization. Their suggested prior distribution for the treatment effect parameter is a normal distribution with mean zero and standard deviation~2.82. For the heterogeneity parameter~$\tau$, a vague prior is not suggested, especially for the meta-analysis of few studies \citep{RoeverEtAl2021}. \citet{Friede2017} showed that weakly informative priors for $\tau$ display desirable long-run frequentist properties in the meta-analysis of few studies. Their suggestion includes a half-normal prior with scale~0.5 or~1.0 for the meta-analysis of binary outcomes. Other commonly used distributions for~$\tau$ include uniform and Cauchy distributions.

In an MBMA, it is common to use vague priors for the slope parameters~$\alpha$ (in linear or log-linear models) or~$\Emax$ ($\Emax$ or sigmoidal~$\Emax$ models). For the $\EDfifty$~parameter, however the situation is different, since it enters the model non-linearly. In the frequentist framework, it is common to impose bounds on the parameter range of~$\EDfifty$, e.g., zero as a natural lower bound and some upper bound. In the Bayesian framework, one can use a uniform prior with some prespecified bounds for the $\EDfifty$~parameter. However, uniform priors are influenced heavily by the parametrization of~$\EDfifty$. To overcome this, \citet{Bornkamp2012} suggested the use of functional uniform priors for the $\EDfifty$~parameter in an $\Emax$~model. Functional uniform priors are distributed uniformly on the potential different shapes of the underlying nonlinear dose-response function, thus invariant to reparametrization. An approximation of the functional uniform prior can be obtained by re-scaling~$\EDfifty$ with the maximum available dose~$D$. The resulting prior for the $\EDfifty/D$ is the log-normal distribution with mean~\mbox{$-2.5$} and standard deviation~$1.8$. This prior is available in the \pkg{MetaStan} package. We refer to \citet{Bornkamp2014} and \citet{Guenhan2020shrinkage} for more details about the functional uniform priors and their implementation.

\section{Implementation in Stan} \label{sec:stan}
\proglang{Stan} is a probabilistic programming language employing a modern Markov chain Monte Carlo (MCMC) algorithm (\emph{Hamiltonian Monte Carlo}; \citet{Stan}) that can be used to facilitate Bayesian computations. The package \pkg{rstan} \citep{rstan} provides an interface for \proglang{R} to \proglang{Stan}. One needs to be familiar with the \proglang{Stan} modelling language to be able to fit models using \pkg{rstan}. However, learning \proglang{Stan}'s syntax may prevent some practitioners to benefit from \proglang{Stan}. For this purpose, we have developed the \proglang{R} package \pkg{MetaStan} \citep{MetaStan}, which includes pre-compiled \proglang{Stan} code for the meta-analysis models described in Section~\ref{sec:models}. In this section, we will describe the \proglang{Stan} code implemented in \pkg{MetaStan}. 

The parametrization of a statistical model, for example using a centered parametrization, may affect the performance of an MCMC algorithm in terms of \emph{convergence} or \emph{mixing} \citep{MCMCinPractice}. In the presence of sparse data, a non-centered parametrization instead of a centered parametrization has been suggested by \citet{Betancourt2015} in order to improve MCMC performance within a \proglang{Stan} application. A non-centered parametrization of the statistical model described in Equation~(\ref{eq:PMA}) is given by 
\begin{equation}
      G(\theta_{ij}) \;=\;  \left\{
      \begin{array}{ll}
        \mu_{i} - 0.5 (d + u_{i} \cdot \tau),  &\mbox{for $j=0$ (control arm)} \\
        \mu_{i} + 0.5 (d + u_{i} \cdot \tau), & \mbox{for $j=1$ (treatment arm)} \label{eq:Reparam}
      \end{array} \right.
\end{equation}
where $u_{i}$ follows a standard normal distribution \citep{Guenhan2020Meta}. The difference between Equations~(\ref{eq:PMA}) and~(\ref{eq:Reparam}) is that in the latter, unit variances~($u_{i}$) are initially utilized and subsequently re-scaled. The implied model, however, remains identical.
Based on the data application at hand, one may want to use a centered or a non-centered parametrization; both are available for pairwise meta-analysis, meta-regression, and model-based meta-analysis in \pkg{MetaStan}.

For MBMA, the heterogeneity random-effects are defined based on a multivariate normal distribution for multi-arm trials (see Equation~(\ref{heterVar})). In terms of the efficiency and stability, it is recommended to use a Cholesky decomposition of the covariance matrix \citep{StanManual}. The \proglang{Stan} implementation in \pkg{MetaStan} to calculate heterogeneity random-effects for the multi-arm trials proceeds in three steps as follows:
\begin{enumerate}
\item Calculate the lower triangular matrix $\boldsymbol{L}_{\gamma}$ such that $\boldsymbol{\Sigma}_{\gamma} = \boldsymbol{L}_{\gamma} \boldsymbol{L^\prime}_{\gamma}$
\item Generate random-effects from $\boldsymbol{u_{i}} = \{u_{i,1}, u_{i,2}, \dots u_{i,T-1} \}$, where each element of $\boldsymbol{u_{i}}$ is drawn from $ \normaldistn(0, 1)$.
\item Multiply $\boldsymbol{u_{i}}$ with $\boldsymbol{L}_{\gamma}$.
\end{enumerate} 

The \proglang{Stan} code for the pairwise meta-analysis and model-based meta-analysis are provided as separate files in the online supplementary document.

\section{Using MetaStan} \label{sec:usage}

\subsection{Pairwise meta-analysis} \label{sec:PMA2}
As an illustrative application, we use a meta-analysis reported by \citet{Boucher2016}. The data set includes six placebo-controlled trials, in which topiramate was investigated for migraine prophylaxis. As one of the outcomes, the occurence of \emph{paresthesia} (abnormal sensation of the skin) was considered. One of the aspects in which the studies differed was the study duration; three lasted for only 8--12~weeks, while tree had a longer follow-up of 18~weeks. The migraine dataset is summarized in Table~\ref{tab:Data1}.

\begin{table}[b]
\centering
\caption{\label{tab:Data1}Data on paresthesia probabilities from the meta-analysis in migraine prophylaxis reported by \citet{Boucher2016}. The numbers of paresthesia occurences and total sample sizes in the control and experimental arm are given for each trial. The six studies differed in terms of their follow-up durations (\emph{short} (8--12~weeks) vs.\ \emph{long} (18~weeks)).}
\begin{tabular}{clcc@{$\,/\,$}cc@{$\,/\,$}c}
  \toprule
  & & duration & \multicolumn{2}{c}{control}   & \multicolumn{2}{c}{treatment} \\
  \cmidrule(rl){4-5} \cmidrule(rl){6-7} 
  $i$ & publication & (weeks) & events  & total & events & total\\ \midrule
  1 &  \citet{Edwards2000}     & 8--12 & 4 & \phantom{0}73 &           63 &           140\\ 
  2 &  \citet{Storey2001}      & 8--12 & 8 &           116 &           53 &           113\\ 
  3 &  \citet{Brandes2004}     & 18 & 9 &           143 &           81 &           144\\ 
  4 &  \citet{Diener2001}      & 18 & 5 &           113 &           57 &           117\\ 
  5 &  \citet{Silberstein2004} & 18 & 4 & \phantom{0}21 &           13 & \phantom{0}19\\ 
  6 &  \citet{Silberstein2006} & 8--12 & 4 & \phantom{0}15 & \phantom{0}9 & \phantom{0}15\\ 
  \bottomrule
\end{tabular}
\end{table}

To analyze the topiramate data, we can load it from the package:
\begin{CodeChunk}
\begin{CodeInput}
R> data("dat.Boucher2016.pairwise")
\end{CodeInput}
\begin{CodeOutput}
               study duration r1  n1 r2  n2
1     Edwards (2000)    short  4  73 63 140
2      Storey (2001)    short  8 116 53 113
3     Brandes (2004)     long  9 143 81 144
4      Diener (2004)     long  5 113 57 117
5 Silberstein (2004)     long  4  21 13  19
6 Silberstein (2006)    short  4  15  9  15
\end{CodeOutput}
\end{CodeChunk}
The dataset first of all is in a \emph{one-study-per-row} format. In order to use the \code{meta\_stan()} function to run the analysis, we need to convert the dataset into a \emph{one-arm-per-row} format. The \code{create\_MetaStan\_dat()} function can be used to perform the conversion:
\begin{CodeChunk}
\begin{CodeInput}
R> dat.topi <- create_MetaStan_dat(dat = dat.Boucher2016.pairwise,
+    armVars = c(responders="r", sampleSize="n"))
\end{CodeInput}
\end{CodeChunk}
The ``\code{armVars}'' argument, a character vector, here is used to indicate the naming of variables in the original, as well as the resulting data set. See Table~\ref{tab:create} for the required variable names based on the scale of the outcome. The columns \code{r1} and \code{r2} denote the numbers of events in the treatment and control groups; the columns \code{n1} and \code{n2} refer to total sample sizes in the migraine dataset. Thus, we use the characters \code{r} and \code{n} to specify the argument \code{armVars} in the above command.
The result then is a \code{list} object containing the data in \emph{one-arm-per-row} format; we may check out the first few lines:
\begin{CodeChunk}
\begin{CodeInput}
> head(dat.topi[[1]])
\end{CodeInput}
\begin{CodeOutput}
  study responders sampleSize na
1     1          4         15  2
2     1          9         15  2
3     2          8        116  2
4     2         53        113  2
5     3          5        113  2
6     3         57        117  2
\end{CodeOutput}
\end{CodeChunk}

\begin{table}[b]
\centering
\caption{Supported endpoints, likelihood and link functions in the \pkg{MetaStan} package. Also, the required column names of the argument \code{armVars} of \code{create\_MetaStan\_dat} are provided.}
\label{tab:create}
\begin{tabular}{llll}
  \toprule
  endpoint & likelihood & link & required columns \\
  \midrule
  dichotomous & binomial & logit    & \code{responders}, \code{sampleSize} \\
  continuous  & normal   & identity & \code{mean}, \code{std.err} \\
  count       & Poisson  & log      & \code{count}, \code{exposure} \\
  \bottomrule
\end{tabular}
\end{table}

The main function to run the pairwise meta-analyses and meta-regressions is called \code{meta\_stan()}; in order to use it, several arguments need to be supplied. The dataset and the likelihood function must be specified. Three likelihood functions (\code{binomial}, \code{normal}, and \code{poisson}) are supported. Random effects are included in the model by default, a common-effect model is used when specifying \code{re = FALSE}. As discussed in Section~\ref{sec:stan}, the choice of parametrization affects MCMC performance. Since the topiramate dataset includes only six studies, a relatively sparse dataset, we use the non-centered parametrization by specifying \code{ncp = TRUE}. Furthermore, the prior distributions for the parameters~$\mu$, $\theta$, and~$\tau$ can be specified. For~$\mu$ and~$\theta$, only normal priors are available. The mean and standard deviation of the normal prior can be specified using the \code{mu\_prior} and \code{theta\_prior} arguments. 
The default parametrization~(\ref{eq:PMA}) may be switched to the alternative~(\ref{eq:PMA2}) using the \code{param}~argument.
Options for the prior distribution for~$\tau$ include half-normal, Cauchy, and uniform distributions. Lastly, the arguments for the MCMC can be specified, which include the number of chains (\code{chains}), the warmup (\code{warmup}) and the total  (\code{iter}) number of iterations. Using the dataset converted into a one-arm-per-row format, we may now execute the analysis using the following command:
\begin{CodeChunk}
\begin{CodeInput}
R> ma.topi <- meta_stan(data=dat.topi, likelihood = "binomial", re=TRUE, 
+    ncp=TRUE, mu_prior=c(0,10), theta_prior=c(0,2.5), tau_prior=0.5,
+    tau_prior_dist="half-normal", chains=4, iter=4000, warmup=2000)
\end{CodeInput}
\end{CodeChunk}

A quick summary of the results can be obtained using the \code{print()} command:
\begin{CodeChunk}
\begin{CodeInput}
R> ma.topi
\end{CodeInput}
\begin{CodeOutput}
Meta-analysis using MetaStan 

Maximum Rhat: 1 
Minimum Effective Sample Size: 3200 

mu prior: Normal(0,10)
theta prior: Normal(0,2.5)
tau prior:half-normal(0.5)

Treatment effect (theta) estimates
 mean  2.5%   50% 97.5% 
 2.65  2.15  2.66  3.14 

Heterogeneity stdev (tau)
     Mean Lower  50% Upper
[1,] 0.26     0 0.21  0.66
\end{CodeOutput}
\end{CodeChunk}

Since \pkg{MetaStan} uses MCMC to explore posterior distributions, it is crucial to investigate convergence diagnostics before reporting the posterior estimates. In the output, some convergence diagnostics are reported at the top, more specifically, the maximum value of the Gelman-Rubin~$\hat{R}$ statistic and the minimum number of effective sample sizes across estimated parameters. Smaller $\hat{R}$~values (closer to~1) and greater effective sample sizes are more desirable. Note that \pkg{rstan} provides many more convergence diagnostics tools and one can obtain them by working with the \code{sampling()} output (as returned in the \code{\$fit} element of the output; see also below). We refer to Section~16 in Stan Reference Manual \citep{StanRef} for more details on the convergence diagnostics. Also see \citet{Sorensen2016} for a gentle introduction to \proglang{Stan}.

Eventually, the prior distributions as well as summary statistics of the posterior distributions the treatment effect~($\theta$) and heterogeneity standard deviation~($\tau$) are given in the output.
In order to visualize the dataset and the summary estimates, the \code{forest_plot()} function may be utilized:
\begin{CodeChunk}
\begin{CodeInput}
R> forest_plot(ma.topi, xlab="log-OR", labels=dat.Boucher2016.pairwise$study)
\end{CodeInput}
\end{CodeChunk}

\begin{figure}[htb]
\centering
\includegraphics[width=1\textwidth]{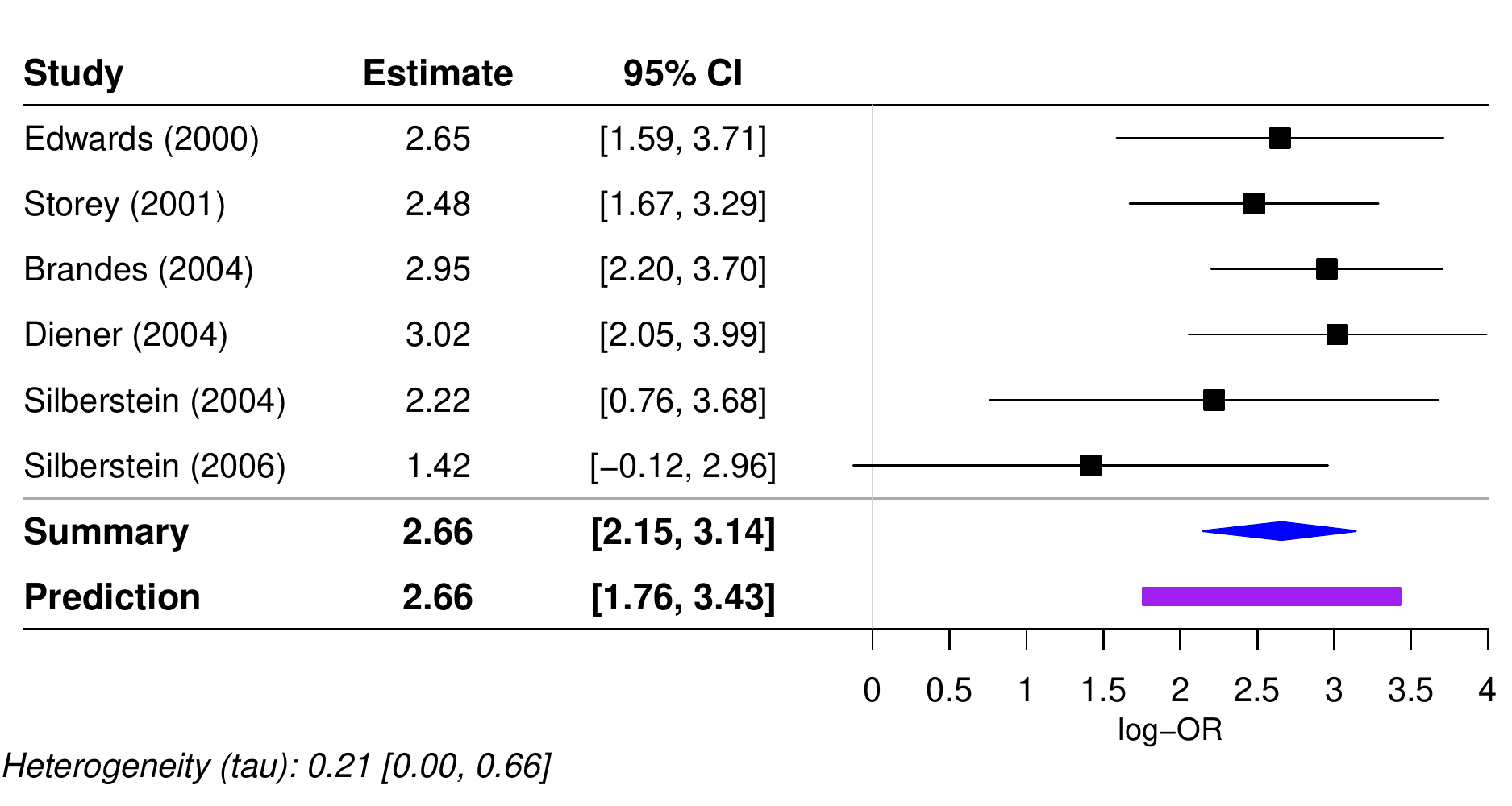}
\caption{A forest plot obtained by the \code{forest\_plot()} function. The meta-analysis is the topiramate example conducted by \citet{Boucher2016}. \label{fig:forest}}
\end{figure}

The resulting forest plot is shown in Figure~\ref{fig:forest}. In the upper section, log odds ratio estimates along with 95\% confidence intervals for each study are shown. In the bottom lines, the treatment effect estimate~($\theta$) and prediction of a new study~($\theta^\star$) along with 95\% credible intervals are shown. The optional argument \code{labels} is used to specify publication names to be displayed in the forest plot. The \code{forest\_plot()} function can be customized by changing axis labels, colors etc. See \code{help(forest\_plot)} in \proglang{R} for all the options.

The \code{meta\_stan()} function internally utilizes the \pkg{rstan} package's \code{sampling()} function to obtain posterior distributions. The output of \code{meta\_stan()} is a \code{list}~object of values including the output from the \code{sampling()} function and the dataset used in the analysis. The output obtained from the \code{sampling()} function can be accessed via \code{ma.topi\$fit}, which yields the original plain MCMC samples. As an example, we can plot the posterior distribution of~$\theta$ as an histogram by calling
\begin{CodeChunk}
\begin{CodeInput}
R> posterior <- as.data.frame(rstan::extract(ma.topi$fit, pars="theta"))
R> hist(posterior$theta)
\end{CodeInput}
\end{CodeChunk}

\subsection{Meta-regression} \label{sec:metareg}
We can fit a meta-regression model using the \code{meta\_stan} function by specifying \code{mreg = TRUE} and including the covariate information to the \code{cov} argument. When a single trial-level covariate is included to the model, the \code{cov} argument accepts a numerical vector. For the case of multiple covariates in the model, a numerical matrix with columns corresponding to different covariates needs to be used. Also, the \code{beta\_prior} arguments specify the mean and standard deviation parameter of the normal prior for~$\beta$.
\begin{CodeChunk}
\begin{CodeInput}
R> mreg.topi <- meta_stan(data=dat.topi, likelihood="binomial",
+    mu_prior=c(0,10), theta_prior=c(0,2.5), tau_prior=0.5, 
+    tau_prior_dist="half-normal", mreg=TRUE, 
+    cov=as.numeric(dat.Boucher2016.pairwise$duration=="long"),
+    beta_prior=c(0,100))
\end{CodeInput}
\end{CodeChunk}
In this example, we included the follow-up duration of each trial (0: short (8--12 weeks); 1: long (18 weeks)) as a covariate in the model. The posterior median estimate of $\beta$ is 0.54 with 95\% credible interval (-0.44, 1.58). 
In the topiramate example, the \emph{long} studies tend to be associated with slightly greater effects (as indicated by the positive regression coefficient), however, including study duration as a covariate does not convincingly explain the source of the heterogeneity between trials, as the uncertainty in the associated $\beta$~estimate is large, and the \emph{short} and \emph{long} studies still exhibit similar effects.

\subsection{Model-based meta-analysis} \label{sec:MBM3}
The migraine prophylaxis data set due to \citet{Boucher2016} in fact contains more information than considered so far; besides the common dosage of 200~mg (and the ``control'' group of 0~mg), some of the trials investigated one or several further doses in additional study arms.
The complete data are shown in Table~\ref{tab:Data2}.

\begin{table}[b]
\centering
\caption{\label{tab:Data2}The extended data set on paresthesia probabilities from the meta-analysis in migraine prophylaxis conducted by \citet{Boucher2016} (see also Table~\ref{tab:Data1}). For each dose considered in a study arm, the number of paresthesia events and the sample sizes are given.}
\begin{tabular}{llcc@{$\,/\,$}cc@{$\,/\,$}cc@{$\,/\,$}cc@{$\,/\,$}c}
  \toprule
   & & & \multicolumn{8}{c}{dose (mg)} \\
  	\cmidrule(rl){4-11}   
  $i$ & publication             & arms & \multicolumn{2}{c}{0} & \multicolumn{2}{c}{50} & \multicolumn{2}{c}{100} & \multicolumn{2}{c}{200} \\ 
  \midrule
  1   & \citet{Edwards2000}     & 2 & 4 & \phantom{0}73  & \multicolumn{2}{c}{} & \multicolumn{2}{c}{} &  63 & 140    \\ 
  2   & \citet{Storey2001}      & 4 & 8 & 116 & 43 & 118 & 59 & 126 & 53 & 113  \\ 
  3   & \citet{Brandes2004}     & 3 & 9 & 143 & \multicolumn{2}{c}{} & 77 & 141 & 81 & 144  \\ 
  4   & \citet{Diener2001}      & 4 & 5 & 113 & 40 & 117 & 59 & 119 & 57 & 117 \\ 
  5   & \citet{Silberstein2004} & 2 & 4 & \phantom{0}21  & \multicolumn{2}{c}{} & \multicolumn{2}{c}{} & 13 & \phantom{0}19     \\ 
  6   & \citet{Silberstein2006} & 2 & 4 & \phantom{0}15  & \multicolumn{2}{c}{} & \multicolumn{2}{c}{} & \phantom{0}9 & \phantom{0}15   \\ 
   \bottomrule
\end{tabular}
\end{table}

The complete dataset is included in the \pkg{MetaStan} package. After loading the data, as before, we need to convert the dataset to a \emph{one-row-per-arm} format using the \code{create\_MetaStan\_dat} function as follows: 
\begin{CodeChunk}
\begin{CodeInput}
R> data("dat.Boucher2016")
R> dat.topi02 <- create_MetaStan_dat(dat=dat.Boucher2016,
+    armVars=c(dose="d", responders="r", sampleSize="n"),
+    nArmsVar="nd")
\end{CodeInput}
\end{CodeChunk}
For the model-based meta-analysis, we need to provide additional variables to the \code{armVars} argument of the \code{create\_MetaStan\_dat()} function, namely the doses (\code{d}) and the number of doses in each trial (\code{nd}).

\begin{figure}[t]
\centering
\includegraphics[width=0.8\textwidth]{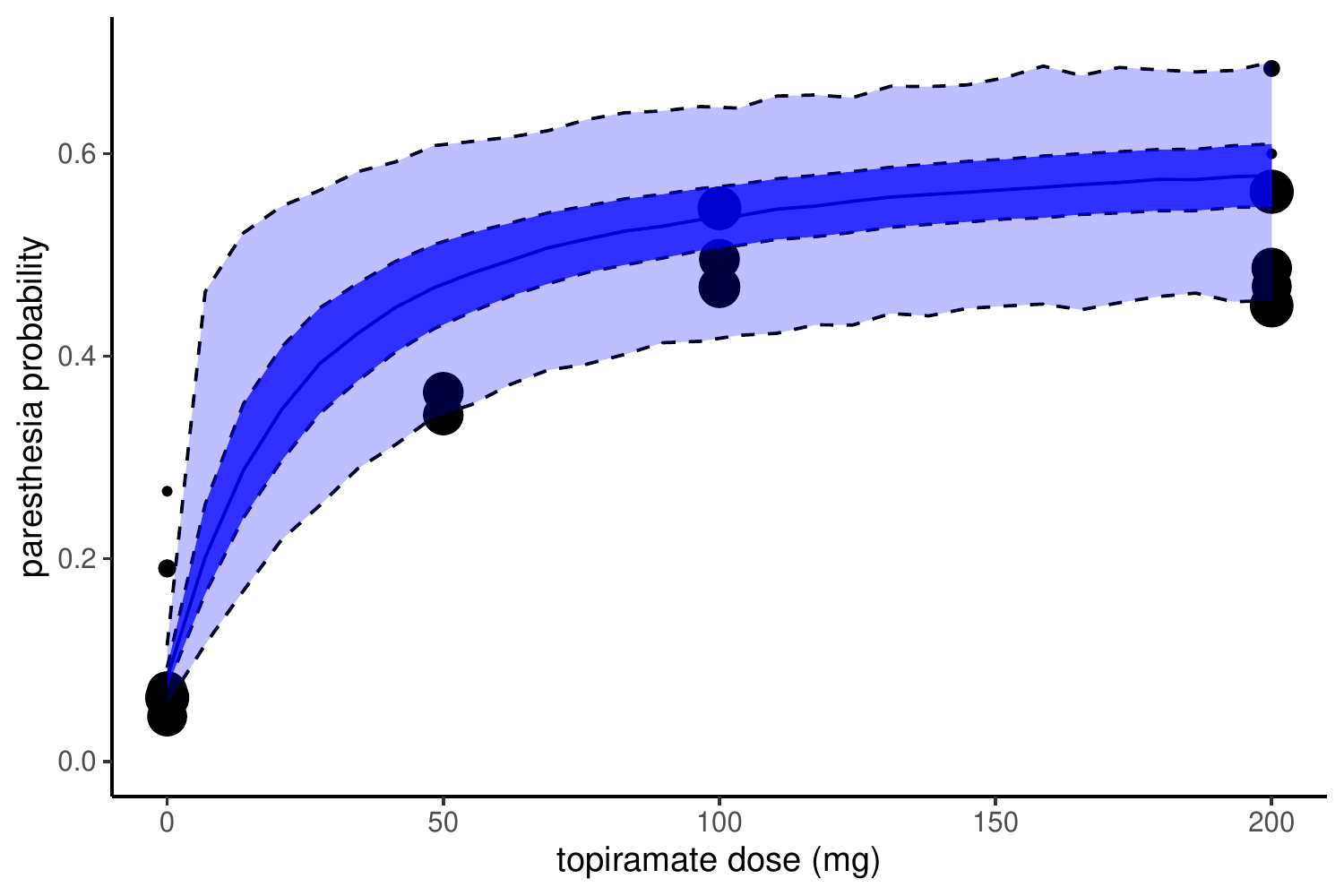}
\caption{\label{fig:doseresponse}A dose-response plot of the migraine dataset using an Emax model. The observed paresthesia probabilities and the estimated dose-response function with pointwise 95\% (light blue) and 50\% (dark blue) credible intervals are displayed. Point sizes are proportional to sample sizes.}
\end{figure}

To fit a model-based meta-analysis, we need to specify a dose-response model using the \code{dose\_response} argument of the \code{MBMA\_stan} function. The available models are \code{linear}, \code{log-linear}, \code{emax} and \code{sigmoidal} as described in Table~\ref{tab:doseresponse}. Similar to \code{meta\_stan}, the arguments \code{data} and \code{likelihood} are required, and prior distributions for the parameters and MCMC settings are optional. See the ``\code{MBMA\_stan()}'' online help for the options and default settings.
\begin{CodeChunk}
\begin{CodeInput}
R> MBMA.emax <- MBMA_stan(data=dat.topi02, likelihood="binomial", 
+    dose_response="emax", Emax_prior=c(0, 10), ED50_prior_dist="functional", 
+    tau_prior_dist="half-normal", tau_prior=0.5)
\end{CodeInput}
\end{CodeChunk}
A quick summary of the output may again be obtained using the \code{print()} command:
\begin{CodeChunk}
\begin{CodeInput}
R> print(MBMA.emax)
\end{CodeInput}
\begin{CodeOutput}
 Model-based meta-analysis using MetaStan 
 
 Maximum Rhat: 1 
 Minimum Effective Sample Size: 1400 
 
 mu prior: Normal(0,10)
 alpha prior: Normal(0,10)
 ED50 prior:functional(-2.5,1.8)
 
 tau prior:half-normal(0.5)
 
 Dose-response function = emax 
 
 Emax estimates
  mean  2.5%   50% 97.5% 
  2.92  2.41  2.91  3.46 
 
 ED50 estimates
  mean  2.5%   50% 97.5% 
 14.00  1.46 13.07 32.09 
 
 [1] 0.00 0.12 0.37
\end{CodeOutput}
\end{CodeChunk}
In this example, we can conclude that the default number of iterations was adequate based on~$\hat{R}$ and the effective sample size. The estimated dose-response function alongside with the observed paresthesia probabilities can be illustrated using the \code{plot} function. For some additional customization, we can use the \pkg{ggplot2} package:
\begin{CodeChunk}
\begin{CodeInput}
R> plot(MBMA.emax) + ggplot2::xlab("topiramate dose (mg)") +
+    ggplot2::ylab("paresthesia probability")
\end{CodeInput}
\end{CodeChunk}
Figure~\ref{fig:doseresponse} shows the dose-response plot of the migraine dataset using the Emax model. The observed paresthesia probabilities and the estimated dose-response function with pointwise 95\% (light blue) and 50\% (dark blue) credible intervals are also displayed in the plot. The point sizes are proportional to sample sizes.

\section{Conclusions} \label{sec:discuss}
In the paper, we present a general overview of the \pkg{MetaStan} \proglang{R} package for fitting pairwise and model-based meta analyses using \proglang{Stan}. \proglang{Stan} provides a general toolbox for Bayesian parameter estimation, however, adapting it for specific purposes is often cumbersome. While software for fitting basic meta-analysis models is available, the \pkg{MetaStan} package also provides generalizations that are useful in practice, namely, one-stage methods for various outcomes, as well as extensions to meta-regression and model-based meta-analysis. \pkg{MetaStan} eliminates the burden for the user to worry about implementation, parametrization details and \proglang{Stan} syntax, and allows to focus on the analysis and interpretation.

The models implemented in \pkg{MetaStan} use a parametrization in which relative treatment effects are assumed to be exchangeable. These models are called contrast-based models. Instead, in an arm-based model, absolute treatment effects are assumed to be exchangeable \citep{Dias2015, White2019}. A possible extension of \pkg{MetaStan} is to include arm-based models for the pairwise and model-based meta-analysis. Another possible extension of \pkg{MetaStan} is to include network meta-analysis model \citep{Guenhan2018NMA}, which enables the analysis of multiple treatments and/or multi-arm trials in the dataset. 

\section*{Acknowledgements}
We would like to thank the Stan Development Team for creating \proglang{Stan}, which is a very powerful tool for Bayesian inference.

\bibliography{references}

\end{document}